\documentclass[pra,twocolumn,showpacs,superscriptaddress]{revtex4-1}
\usepackage{amsmath}
\usepackage{amssymb}
\usepackage{graphicx}
\usepackage{color}
\def\ket#1{\left|#1\right\rangle}

\begin{document}

\title{Measurement of a topological edge invariant in a microwave network}

\author{Wenchao Hu}

\affiliation{Centre for Disruptive Photonic Technologies, Nanyang Technological University, Singapore 637371, Singapore}

\author{Jason C.~Pillay}

\affiliation{Division of Physics and Applied Physics, School of Physical and Mathematical Sciences, Nanyang Technological University, Singapore 637371, Singapore}

\author{Kan Wu}

\affiliation{State Key Laboratory of Advanced Optical Communication Systems and Networks, Department of Electronic Engineering, Shanghai Jiao Tong University, Shanghai 200240, China}

\author{Michael Pasek}

\affiliation{Division of Physics and Applied Physics, School of Physical and Mathematical Sciences, Nanyang Technological University, Singapore 637371, Singapore}

\author{Perry Ping Shum}

\affiliation{Centre for Disruptive Photonic Technologies, Nanyang Technological University, Singapore 637371, Singapore}

\author{Y.~D.~Chong}
\email{yidong@ntu.edu.sg}

\affiliation{Centre for Disruptive Photonic Technologies, Nanyang Technological University, Singapore 637371, Singapore}

\affiliation{Division of Physics and Applied Physics, School of Physical and Mathematical Sciences, Nanyang Technological University, Singapore 637371, Singapore}

\begin{abstract}
  We report on the measurement of topological invariants in an
  electromagnetic topological insulator analog formed by a microwave
  network, consisting of the winding numbers of scattering matrix
  eigenvalues.  The experiment can be regarded as a variant of a
  topological pump, with non-zero winding implying the existence of
  topological edge states.  In microwave networks, unlike most other
  systems exhibiting topological insulator physics, the winding can be
  directly observed.  The effects of loss on the experimental results,
  and on the topological edge states, is discussed.
\end{abstract}

\pacs{42.60.Da, 42.70.Qs, 73.43.-f}

\maketitle

\section{Introduction}

Topological insulators are phases of matter that are ``topologically
distinct'' from conventional insulators, meaning that their electronic
bandstructures cannot be deformed into conventional bandstructures
without closing bandgaps.  They have many striking physical
properties, the most notable being the existence of ``topological edge
states'' which, in two dimensional (2D) topological insulators, have
the unique property of being protected against backscattering from
impurities.  Topologically nontrivial bandstructures were originally
discovered in condensed matter systems, first in the quantum Hall (QH)
effect \cite{MStone} and later in materials with strong spin-orbit
coupling \cite{Moore}.  Recently, the concept has been extended to
photonics, with coherent classical electromagnetic fields taking the
place of electron wavefunctions
\cite{Raghu1,Raghu2,Wang1,Wang2,WenjieChen,Rechtsman,hafezi,hafezi2,Fan,LeHur,LeHur1,Khanikaev}.
Such ``topological photonics'' devices have been realized with
microwave-scale magnetic photonic crystals
\cite{Raghu1,Raghu2,Wang1,Wang2} and meta-atom structures
\cite{WenjieChen}; and, at optical and infrared frequencies, with
waveguide lattices \cite{Rechtsman} and resonator lattices
\cite{hafezi,hafezi2}.  There have also been theoretical proposals
based on modulated photonic crystal resonances \cite{Fan}, circuit QED
systems \cite{LeHur,LeHur1}, metamaterial photonic crystals
\cite{Khanikaev}, etc.  The key feature of these devices is the
existence of topologically protected electromagnetic edge states,
which may have technological promise for waveguides that are robust
against disorder.  Topological photonics may also prove useful for
studying aspects of topological insulator physics which are difficult
or impossible to probe in the condensed-matter context, such as the
effects of nonlinearity \cite{solitons}.

In condensed matter systems, the most important physical consequence
of topologically nontrivial bandstructures is that they cause certain
macroscopic transport properties to be precisely quantized.  Most
famously, in the integer QH effect, the Hall conductance is quantized
to integer multiples of the inverse von Klitzing constant to nearly
one part in $10^9$ \cite{klitzing}.  An influential explanation for
this was supplied by Thouless \textit{et al.}~\cite{TKNN}, who showed
using linear response theory that the QH conductance is tied to the
Chern numbers of the bands, which are ``topological invariants''
restricted to integer values.  In photonics, however, there is no
direct analog of the Hall conductance or similar linear response-based
quantity, due to the absence of a fermionic ground state
\cite{Raghu1,Raghu2}; this is why edge propagation measurements have
served almost exclusively as the signature for topologically
nontrivial photonic bandstructures.

In this paper, we present experimental measurements of a topological
edge invariant in a microwave network.  The invariant consists of the
integer winding numbers of scattering matrix ($S$ matrix) eigenvalues,
measured at the edges of the sample.  The experiment is a variant of
the ``topological pump'', a thought experiment invented by Laughlin to
explain the QH effect in topological terms---in a way that does
\textit{not} rely on linear response \cite{Laughlin, Halperin}.  In Laughlin's
original setup, the Hall conductance is described in terms of the
adiabatic ``pumping'' of single-electron wavefunctions across the
surface of a cylinder; due to the cylindrical geometry, the pumping
can be expressed as the effect of a gauge transformation, whereupon
the robustness of the QH effect follows as a result of the exactness
of gauge invariance.  The topological pump can also be formulated in
terms of scattering processes \cite{Brouwer0,Brouwer,Fulga},
which is a particularly natural approach for describing
electromagnetic systems \cite{pasek}.  As described below, the present
experiment is based on this scattering formulation.  It is remarkable
that, even though the topological pump is typically regarded as a
``thought experiment'', it can actually be implemented in
electromagnetic systems.  Recently, topological pumps have been
demonstrated in 1D quasicrystalline arrays of optical waveguides
\cite{kraus,verbin}, and a related scheme based on a ring of coupled
optical resonators has been proposed by Hafezi \cite{hafezi3}.
Compared to these previous works, our microwave experiment is
noteworthy in that the measurements are not limited to intensities,
but include full phase information, since the complex $S$ matrix can
be determined with a microwave network analyzer; this allows the
winding process to be directly observed.  The difference between
topologically trivial and nontrivial behaviors is also clearly
observable in this system.  The setup also provides a platform for
studying the topological properties of ``Floquet'' or ``quasi-energy''
bandstructures, a topic of current theoretical interest
\cite{Oka,Inoue,Demler0,Demler,Lindner,Gu,Levin}.

The remainder of this paper is organized as follows: Section II
reviews the theory of network bandstructures and topological pumps;
Section III describes the experimental setup and results; finally,
Section IV discusses the implications of the experiment, including how
we can understand the effects of losses on the topological edge
invariant and the network bandstructure.

\section{Bandstructures and topological pumps in networks}

\subsection{Network bandstructures}
\label{Network bandstructures}

Our realization of an electromagnetic topological insulator is a
network implemented with microwave components.  Such a system is best
understood using the framework of a ``network model''
\cite{ChalkerCo}, which differs in several respects from the
tight-binding Hamiltonian models familiar to most readers.  This
section describes our network model and its topological properties.  A
more detailed discussion may be found in Ref.~\cite{pasek}.

\begin{figure}
  \centering\includegraphics[width=0.46\textwidth]{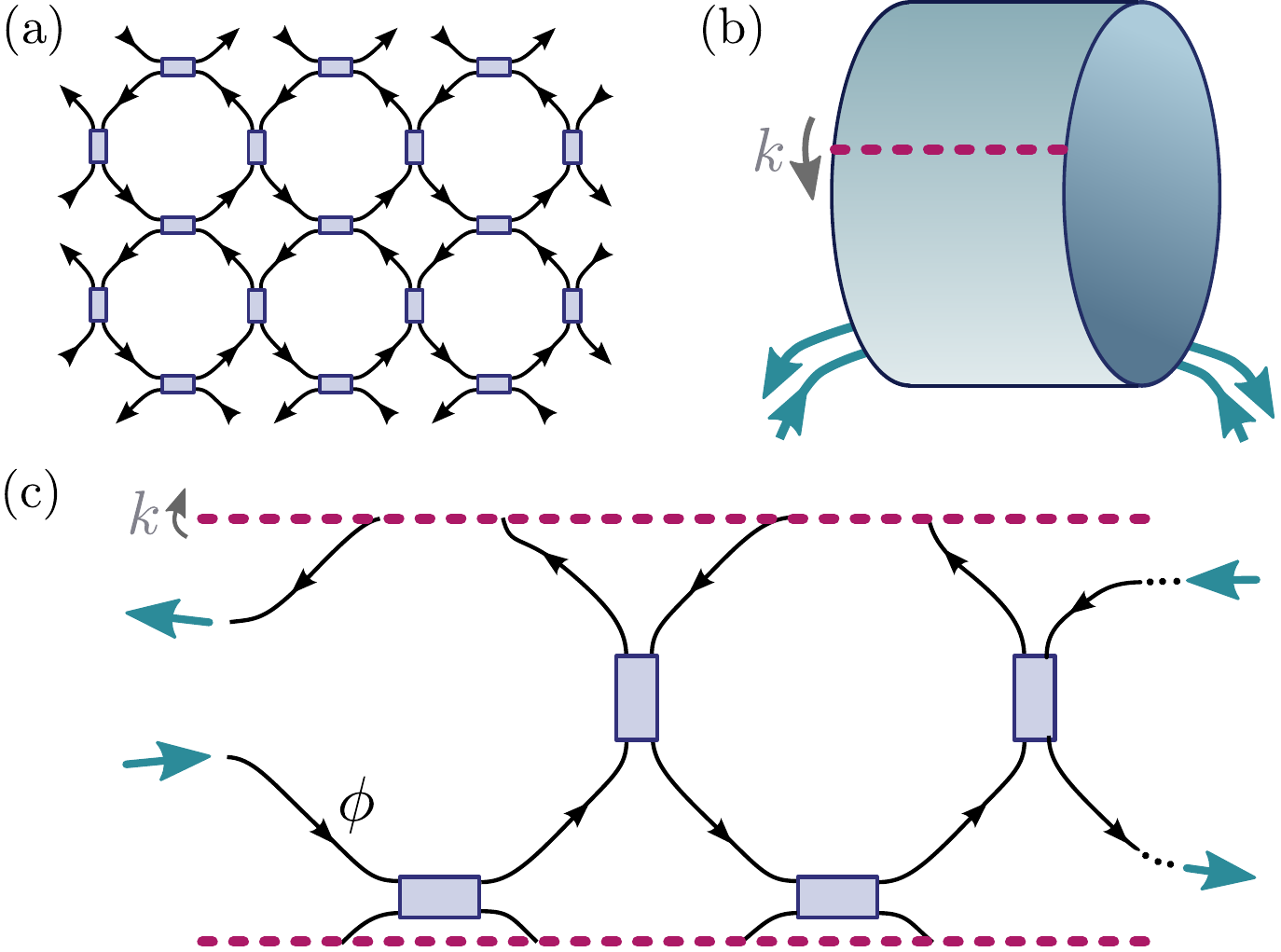}
  \caption{(a) Schematic of a periodic 2D directed network.  Wave
    amplitudes undergo phase delay $\phi$ along each directed link
    (black arrows), and couple through a $2\times2$ unitary matrix at
    each node (blue boxes). (b) In a ``topological pump'' setup, a 2D
    system is rolled into a cylinder; twisted boundary conditions with
    tunable twist angle $k$ are applied, and a transport measurement
    is performed in the axial direction (cyan arrows).  (c)
    Topological pump setup for a network model, with a cylinder
    comprising one unit cell in the azimuthal direction and two cells
    in the axial direction. }
  \label{fig:schematic}
\end{figure}

Network models, which originated in the QH literature as a convenient
way to study disordered QH systems \cite{ChalkerCo} and other types of
topological insulators \cite{Kramer,ryu}, describe directed networks
within which waves can propagate.  We consider the specific periodic
(disorder-free) network shown in Fig.~\ref{fig:schematic}(a), which
consists of unit cells arranged in a 2D square lattice, each cell
containing two nodes and four directed links.  A ``solution'' to the
network is a set of wave amplitudes (complex scalars) defined along
the links, such that (i) traversing each link incurs a phase delay
$\phi$ (which we take to be the same for all links), and (ii) the
amplitudes entering and leaving each node are related by a fixed
$2\times2$ coupling matrix.  Let the complex vector $\ket{\psi}$ be
the collective set of wave amplitudes exiting the links, and
$\ket{\psi'} = e^{-i\phi} \ket{\psi}$ be the amplitudes entering the
links.  The individual node coupling matrices can be composed into a
matrix $U$ that relates the amplitudes entering and leaving the nodes.
These amplitudes must be equal to $\ket{\psi}$ and $\ket{\psi'}$
respectively, so
\begin{equation}
  U \ket{\psi} = e^{-i\phi} \ket{\psi}.
  \label{Ueq}
\end{equation}
For an infinite periodic network, we look for solutions obeying
Bloch's theorem.  Then Eq.~(\ref{Ueq}) reduces to \cite{Liang,pasek}
\begin{equation}
  U_k \ket{\psi_{nk}} = e^{-i\phi_n(k)} \ket{\psi_{nk}},
  \label{Uk}
\end{equation}
where $n$ is the band index, $k$ is the quasimomentum, and $\phi_n(k)$
is the band ``quasi-energy''.  In the absence of gain or loss in the
network, $U_k$ is unitary, and $\phi_n(k)$ is real.  But unlike a
usual band energy, it is an angle variable.

Interestingly, Eq.~(\ref{Uk}) has the same form as the evolution
equation for a lattice with a time-periodic Hamiltonian, where $U_k$
is the evolution operator over one period and $\ket{\psi_{nk}}$ is a
Floquet state of the driven lattice.  Recently, several groups have
proposed a new class of topological insulators called ``Floquet
topological insulators'', where the topological behavior arises from
periodic drives (such as oscillating electric fields).  It has been
shown that a driven system can have a topologically nontrivial
\textit{quasi-energy} bandstructure, even if the undriven
\textit{energy} bandstructure is topologically trivial
\cite{Oka,Inoue,Demler0,Demler,Lindner,Gu,Levin}.

The network model framework is also useful for describing a type of
photonic device proposed and realized by Hafezi \textit{et al.},
consisting of a lattice of ring resonators joined by auxiliary
waveguides \cite{hafezi,hafezi2}.  By engineering incommensurate
modulations into the auxiliary waveguides, those authors induced an
effective tight-binding ``vector potential'' corresponding to a
uniform magnetic field.  The system thus maps onto a QH system, and
exhibits electromagnetic analogs of QH edge states \cite{hafezi2}.
Surprisingly, however, even if the auxiliary waveguides are
symmetrical and commensurate, corresponding to zero ``magnetic flux'',
full-wave simulations show that robust one-way edge states can still
exist \cite{Liang2}.  This is explained by moving beyond the
tight-binding description, and formulating a network model where the
arms of the ring resonators are links \cite{Liang,Liang2,pasek}.  That
network model is equivalent to the one considered in this paper.

In most respects, quasi-energy bandstructures have the same behaviors
as energy bandstructures.  If quasi-energy bands have nonzero Chern
numbers (as calculated from $\ket{\psi_{nk}}$ in the usual way
\cite{TKNN}), topological edge states will exist; this is the case
demonstrated in a recent photonic realization of a Floquet topological
insulator \cite{Rechtsman}.  However, a quasi-energy bandstructure can
also be topologically nontrivial \textit{even if all Chern numbers are
  zero} \cite{Demler}.  This case, which we call the ``anomalous
Floquet insulator'', is possible because $\phi_n(k)$ is an angle
variable; for example, every band can receive +1 to its Chern number
from the band above and -1 from the band below, resulting in zero net
Chern number even though all bandgaps are topologically nontrivial.
For driven systems, Rudner \textit{et al.}~have proposed an
alternative topological bulk invariant which can characterize this
situation \cite{Levin}.  The topological nontriviality of an anomalous
Floquet insulator can also be verified from the presence of
topological edge states, and from its behavior under topological
pumping (as described in the next section) \cite{pasek}.

For the periodic network of Fig.~\ref{fig:schematic}, the quasi-energy
bandstructure can be obtained analytically \cite{Liang}.  It turns out
that the band topology depends on a single parameter $\theta \in
[0,\pi/2]$, which describes the coupling strength at the network
nodes.  The bandgaps close at $\theta = \pi/4$.  For $\theta < \pi/4$,
the system is a topologically trivial conventional insulator; when
$\theta > \pi/4$, the system is an anomalous Floquet insulator
\cite{pasek}.  In the experiment, this allows us to switch easily
between topologically trivial and nontrivial behaviors.


\subsection{Topological pumps}

A topological pump is an experiment which is specially designed to
reveal the topological properties of a 2D lattice.  As shown
schematically in Fig.~\ref{fig:schematic}(b), it consists of rolling a
2D lattice into a cylinder, inducing a phase ``twist'' $k$ in the
azimuthal boundary conditions, and performing a transport measurement
in the axial direction (i.e., at the edges of the cylinder).  In the
original Laughlin thought experiment \cite{Laughlin}, the twist is
implemented by adiabatically threading magnetic flux through the
cylinder, which produces an Aharanov-Bohm phase shift while also
transporting electron wavefunctions in the axial direction.  Threading
one magnetic flux quantum induces a $2\pi$ phase shift; by gauge
invariance, the number of transported electrons must be an integer.
From this, the quantization of the Hall conductance can be derived
\cite{Laughlin}.

An alternative formulation of the topological pump, based on wave
scattering, has been developed by Brouwer and co-workers
\cite{Brouwer0,Brouwer}.  Here, one imagines taking a similar
cylinder, and scattering electron waves (or electromagnetic waves, as
the case may be) off one edge, at an energy (or frequency) which lies
in a bulk bandgap.  As the twist angle $k$ advances by $2\pi$, the
reflection matrix $r$ is measured.  For a sufficiently long cylinder
and/or a sufficiently large bandgap, transmission to the opposite edge
is negligible and $r$ is unitary; its eigenvalues lie on the unit
circle, and their trajectories during a pumping cycle can be
topologically characterized by a winding number.  A nonzero winding
number corresponds to a topologically nontrivial sample.  This is tied
to the existence of topological edge states.  In a topologically
nontrivial system, edge states must occur at certain values of $k$
during the pumping cycle \cite{Halperin}.  According to scattering
theory, each such occurrence induces a $\pi$ scattering phase shift
\cite{Brouwer}.  Such a phase shift can only be guaranteed if the
eigenvalue trajectories have nonzero winding.  This behavior is
``topologically protected'', since weak perturbations deform the
trajectories of the reflection eigenvalues without altering the
winding number, and the only way to remove the winding is to close the
band gap.

Our experiment implements a similar scheme, in the context of a
microwave network, as shown in Fig.~\ref{fig:schematic}(c).  The
tunable twist $k$ is implemented by connecting the links (microwave
cables) at each end of the ``azimuthal axis'' to phase shifters.  We
measure the full scattering parameters along the axis (i.e., the
complex reflection and transmission coefficients for microwave signals
injected into the edges of the cylinder), thus obtaining the $S$
matrix, which is $2\times 2$ for this network geometry.  In the ideal
situation where the network is completely lossless, $S$ is unitary,
and its eigenvalues $\sigma_{\pm}$ lie on the unit circle.  If the
quasi-energy $\phi$ lies in a sufficiently large bandgap, then the
transmission goes to zero, and $\sigma_{\pm}$ reduce to the reflection
coefficients from each edge (this condition can indeed be met, as
discussed in Section \ref{discussion section}).  The winding numbers
of $\sigma_{\pm}$, as the twist angle $k$ is tuned through $2\pi$, are
the desired integer invariants \cite{Brouwer0,Brouwer,Fulga,pasek}.

Several other schemes to directly measure topological invariants have
recently appeared in the literature.  Most prominently, a topological
pump has been realized experimentally using a 1D quasicrystalline
optical waveguide array, which can be mapped formally to a 2D QH
system \cite{kraus,verbin}. A scheme to implement a topological pump
in a 2D system has recently been proposed by Hafezi \cite{hafezi3}.
Here, coupled resonators are arranged in an annulus, with a tunable
phase shift along one column of the annulus; such a configuration is
similar to the Laughlin thought experiment \cite{Laughlin}, and to the
present experiment.  A topological invariant is then inferred from the
transport of resonance peaks in the transmission spectrum along the
edge. This proposal has yet to be realized.

Topological pumps, such as the present experiment and the related
schemes described in the preceding paragraph, are probes of
topological \emph{edge} invariants. There have also been theoretical
proposals to directly measure topological \emph{bulk}
invariants---specifically, Chern numbers---in photonic systems
\cite{Carusotto,Bardyn}. Typically, a bulk invariant is significantly
more challenging to measure than an edge invariant, as it requires
detailed information about the wavefunctions in the bulk, whereas the
latter depends only on the response of the system at a single
frequency or quasi-energy \cite{Brouwer}.  Furthermore, as noted
above, edge invariants can detect topologically nontrivial behaviors
in cases where the band topology cannot be fully characterized by
Chern numbers \cite{pasek}. (A recent experiment has measured a
topological bulk invariant in a specific 1D system \cite{Zeuner}.)

\section{Experimental setup and results}
\label{experiment section}

\begin{figure}
  \centering\includegraphics[width=0.47\textwidth]{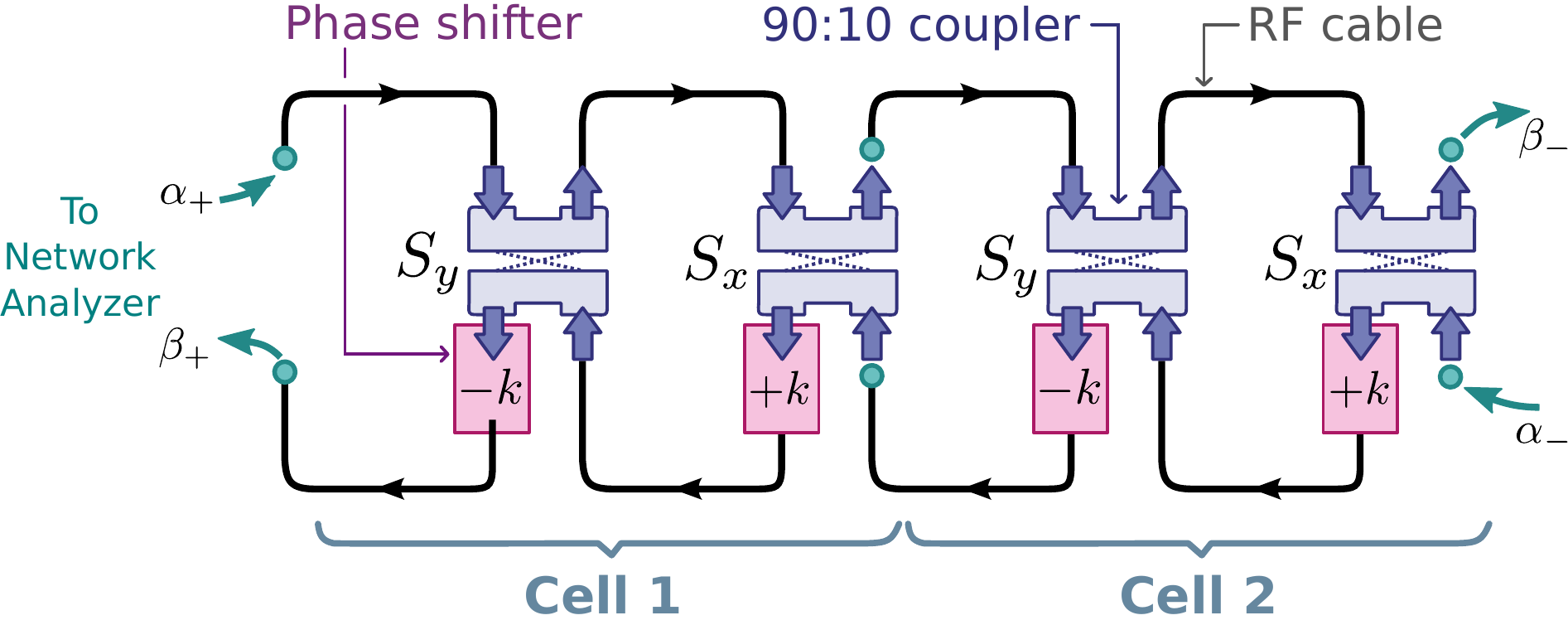}
  \caption{Experimental setup.  Each of the identical units, labeled
    here ``Cell 1'' and ``Cell 2'', corresponds to one cell in the
    topological pump geometry of Fig.~\ref{fig:schematic}.  A pair of
    phase shifters in each cell (pink boxes), with shifts of $+k$ and
    $-k$ respectively, implement the twisted boundary condition.  The
    couplers (blue rods) are depicted in the ``strong coupling''
    configuration; the ``weak coupling'' configuration is achieved by
    swapping each coupler's outputs.  The overall input and output
    amplitudes are $\alpha_\pm$ and $\beta_\pm$; their scattering
    parameters are measured with a network analyzer.  }
  \label{fig:network}
\end{figure}

Our experimental setup is shown in Fig.~\ref{fig:network}.  The
network is divided into identical subunits, each consisting of 4
cables, 2 phase shifters, and 2 couplers.  By comparison with
Fig.~\ref{fig:schematic}(c), each subunit is equivalent to a
``cylinder'' in the topological pump setup that is one cell wide and
one cell long.  Connecting the subunits in series forms a longer
cylinder; Fig.~\ref{fig:network} shows a ``two-cell'' configuration.
The cables are standard low-loss coaxial RF cables of $\sim\!15$ cm
length, and the couplers are four-port single-directional couplers,
with isolators built into each port and coupling ratios of
approximately 90:10 (DTO-2.5/5-10, Shanghai Huaxiang Computer Ltd.).
The two phase shifters in each cell (TKE-90-6SA, Shanghai Huaxiang
Computer Ltd.)  are independently tunable, and set to produce $+k$ and
$-k$ phase shifts respectively.  The operating frequency is chosen to
be 5 GHz, low enough to reduce losses in the various components while
high enough to allow phase shifts in the full range $k \in [0,2\pi]$.
There is one input port and one output port at each end of the
network, connected to a vector network analyzer (Anritsu 37396C).
From this, we measure the $S$ matrix, defined as
\begin{equation}
  S \begin{bmatrix}\alpha_+ \\ \alpha_-
  \end{bmatrix} = 
  \begin{bmatrix}
    \beta_+ \\ \beta_-
  \end{bmatrix},
\end{equation}
where $\{\alpha_\pm, \beta_\pm\}$ are the input and output wave
amplitudes at the two edges, as labeled in Fig.~\ref{fig:network}.

Since the couplers have fixed coupling ratios, we are unable to
continuously vary the network's coupling strength parameter, on which
the bandstructure topology depends.  However, by swapping the order of
each coupler's output ports, we can switch between the two cases of a
topologically trivial and nontrivial bandstructure.  As described in
Section \ref{Network bandstructures} and Ref.~\onlinecite{Liang}, the
strength of each coupler is described by a parameter $\theta \in
[0,\pi/2]$, where $\theta = 0$ corresponds to zero coupling between
adjacent unit cells and $\pi/2$ is complete coupling.  The
bandstructure is topologically trivial for $\theta < \pi/4$, and
nontrivial for $\theta > \pi/4$, independent of all other coupling
matrix parameters.  Our 90:10 couplers thus allow for either $\theta
\approx \tan^{-1}(3) \approx 0.40\pi$ (topologically nontrivial; this
is the configuration shown in Fig.~\ref{fig:network}) or $\theta
\approx \tan^{-1}(1/3) \approx 0.10\pi$ (topologically trivial).  At
the 5 GHz operating frequency, each cable has phase delay $\approx
0.2\pi$, which lies in a bandgap of the quasi-energy bandstructure for
both the strong and weak-coupling cases.  The loss in each cable is
$\approx 0.4$dB.

\begin{figure}
  \centering\includegraphics[width=0.47\textwidth]{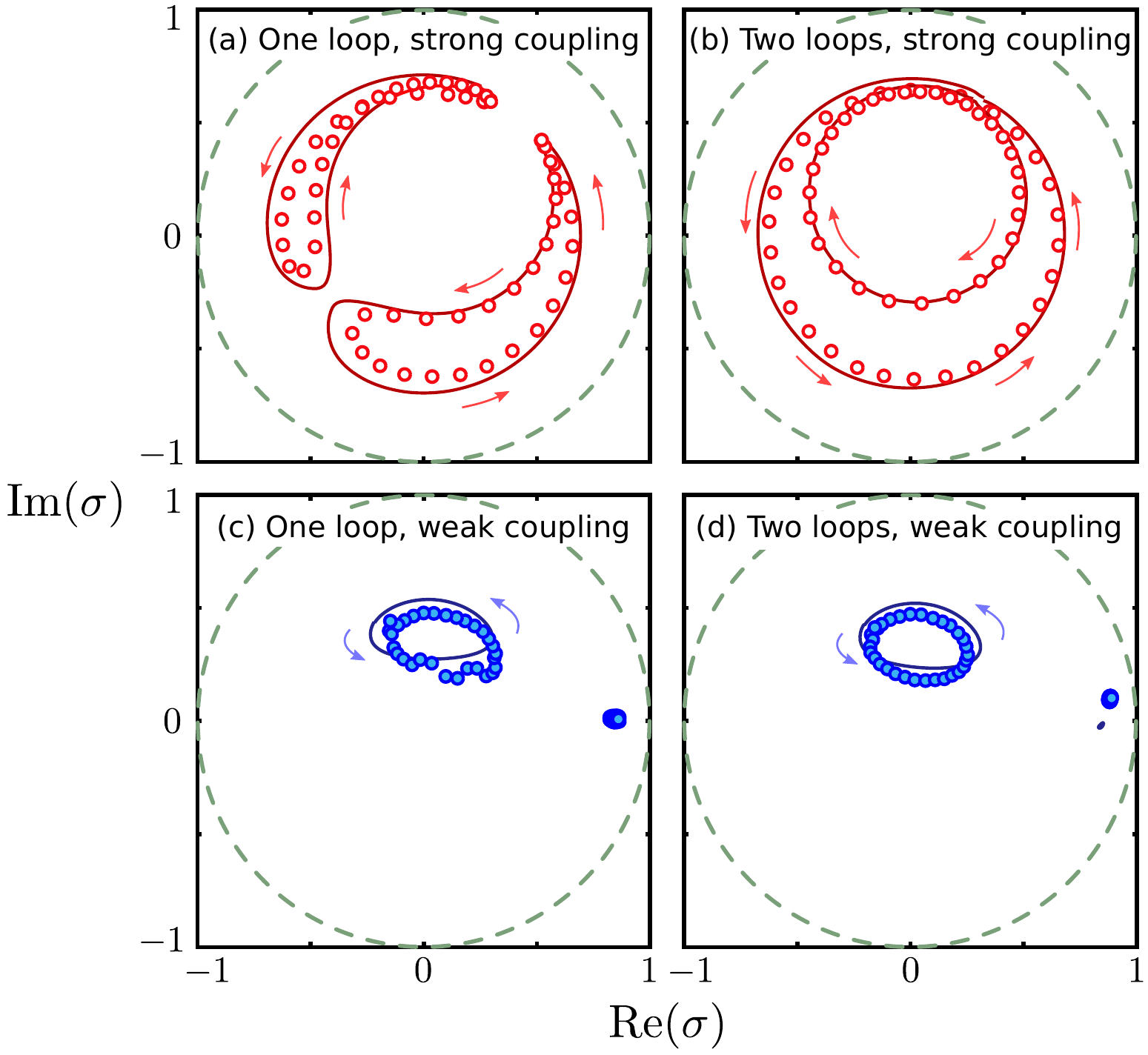}
  \caption{Scattering matrix eigenvalues measured across one cell
    (left) and across two cells in series (right).  Arrows indicate
    the direction of motion with increasing $k$.  Circles show
    experimental data; solid curves show theoretical calculations
    using the scattering parameters measured for each network
    component individually (including losses).  Non-zero winding
    numbers can be observed in the strong-coupling two-cell case.  The
    unit circle is indicated by dashed curves.}
  \label{fig:seigs}
\end{figure}

\begin{figure}
  \centering\includegraphics[width=0.47\textwidth]{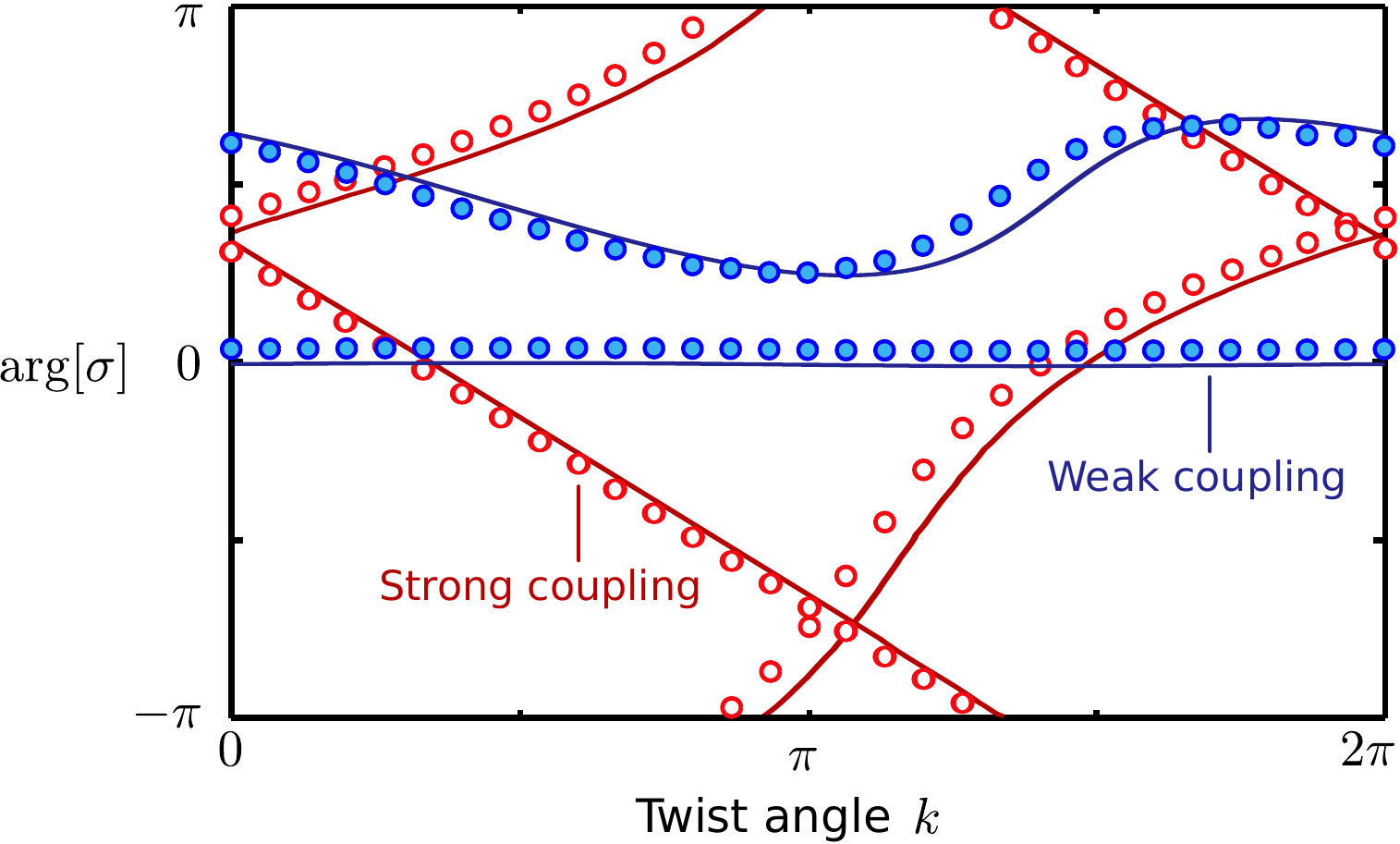}
  \caption{Arguments of the complex $S$ matrix eigenvalues for the
    two-cell network, as the phase shift $k$ is tuned through $2\pi$.
    Empty circles show the strong coupling measurement data, and
    filled circles show the weak coupling data; solid curves show
    theoretical calculations.  For strong coupling, the two
    eigenvalues have winding numbers $\pm 1$, corresponding to the
    bulk bandstructure being topologically nontrivial. }
  \label{fig:winding}
\end{figure}

The measured $S$ matrix eigenvalues are shown in Fig.~\ref{fig:seigs}.
The eigenvalues do not lie on the unit circle, due to losses in the
network which make the $S$ matrix sub-unitary.  Nonetheless, the
eigenvalue trajectories exhibit winding behaviors very similar to the
lossless case.  For the one-cell system under strong coupling, the two
$S$ matrix eigenvalues, $\sigma_{\pm}$, move along distinct closed
trajectories as $k$ is increased through $2\pi$, as shown in
Fig.~\ref{fig:seigs}(a).  Each individual trajectory does not encircle
the origin (i.e.~the winding number is zero), even though the network
bandstructure is in the topologically nontrivial regime.  This is
because topological protection requires the opposite edges in a finite
system to be well separated, so as to have negligible overlap between
the counter-propagating edge states on each edge.  When the separation
is increased by connecting two cells in series, the trajectories
coalesce into a pair of loops with winding numbers $\pm 1$, as shown
in Fig.~\ref{fig:seigs}(b).  By contrast, in the weak coupling regime
of Figs.~\ref{fig:seigs}(c) and \ref{fig:seigs}(d), the eigenvalues
move along separate trajectories without encircling the origin.

The winding numbers can also be visualized by plotting
$\mathrm{arg}[\sigma_\pm]$ against $k$, as in Fig.~\ref{fig:winding}.
It is worth noting that, in the topologically nontrivial regime, the
two $S$ matrix eigenvalues wind in opposite directions.  This
corresponds to the fact that the topological edge states on opposite
edges have opposite group velocities.  In terms of the projected
bandstructure, one branch of edge state have a dispersion curve that
crosses the probed value of $\phi$ from above, and the other from
below; hence, the induced scattering phase shifts have opposite signs.

\section{Discussion}
\label{discussion section}

Our experiment deviates from an ideal topological pump in several
respects.  Firstly, as previously mentioned, in the ideal topological
pump the edges are separated by a large number of unit cells, so that
there is a true ``bulk'', whereas in our experiment there are only two
unit cells.  However, this does not make a significant difference to
the physical interpretation, as the relevant phenomenon---the
emergence of a non-zero winding number---is observed already when
going from the one-cell to the two-cell case, shown in
Figs.~\ref{fig:seigs}(a) and (b).  This arises from the fact that the
system is deep in either the topologically trivial or nontrivial
phases, based on our choices of the coupling strength $\theta$; in the
nontrivial case, each edge state is strongly confined to one unit
cell, with negligible amplitude on the next unit cell.  Indeed,
calculations based on realistic parameters show no significant changes
in the trajectory of the $S$ matrix eigenvalues as the number of cells
is increased beyond two.

When the edges are well-separated, the $S$ matrix reduces to a pair of
reflection coefficients $r_\pm \equiv \beta_\pm / \alpha_\pm$, where
$\{\alpha_\pm, \beta_\pm\}$ are the input/output wave amplitudes
labeled in Fig.~\ref{fig:network}.  This is observed in the small
transmission coefficients in the two unit cell systems: $|t| < 0.06$
in the strong coupling system and $|t| < 0.04$ in the weak coupling
case. Simulations of a larger system (with 5 unit cells) reveal that
the edge states have penetration depths of less than half a unit cell,
as determined by the exponential decay of the wave intensity (see
Appendix A).

Another difference between our experiment and the ideal topological
pump is the presence of loss in all the network components: the
cables, phase shifters, and couplers.  As a result, the eigenvalues of
the $S$ matrix are not strictly constrained to the unit circle, as
seen in Fig.~\ref{fig:seigs}.  Nonetheless, we argue that these
eigenvalue trajectories can be meaningfully linked to the existence or
non-existence of topological edge states, based on the close
relationship between the edge scattering parameters measured in the
experiment and the projected bandstructure of the network.

\begin{figure}
  \centering\includegraphics[width=0.47\textwidth]{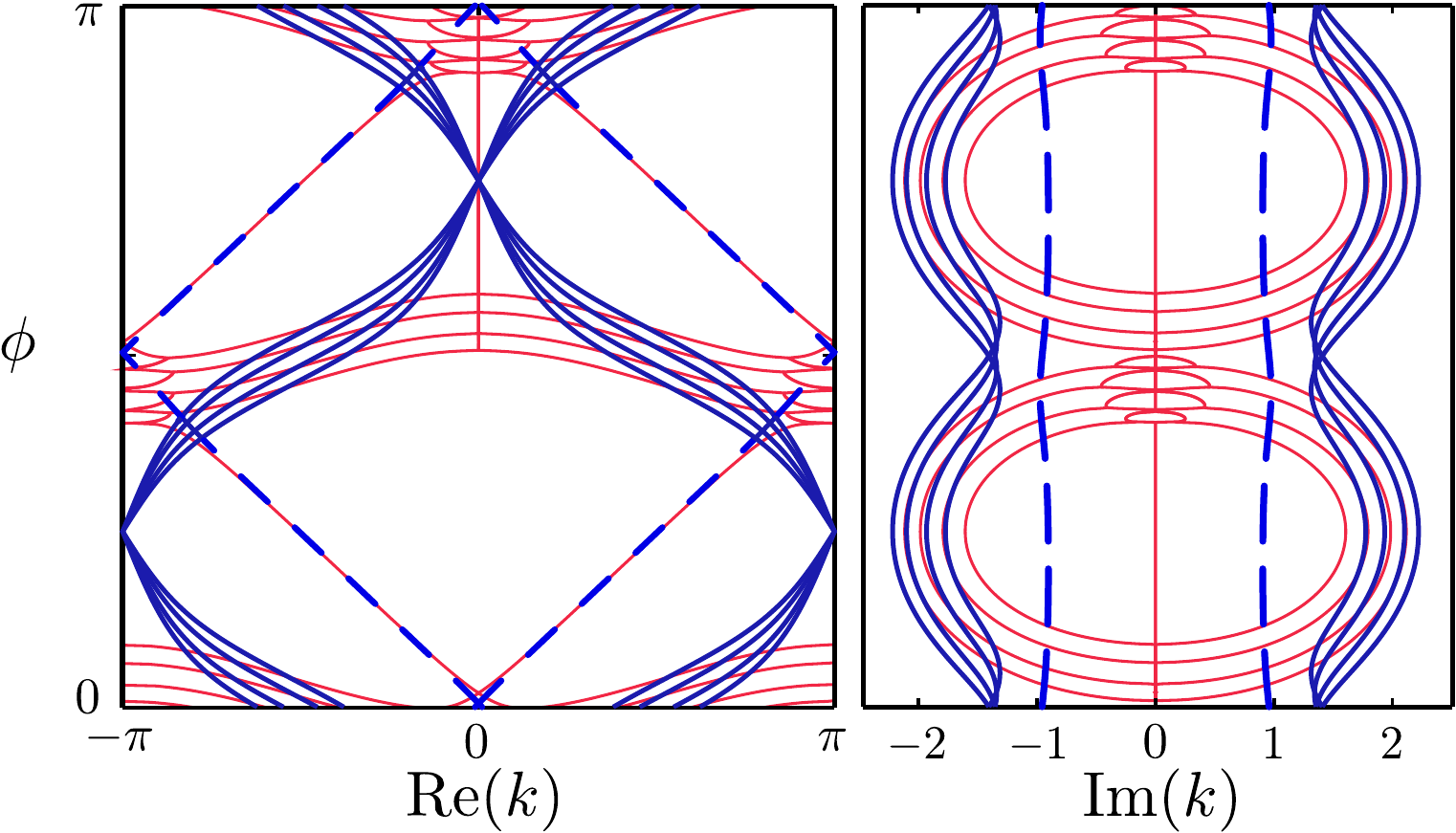}
  \caption{Complex projected bandstructure of a hypothetical lossy
    network (blue lines), consisting of an infinite strip 5 unit cells
    wide.  Each link has real tunable phase delay $\phi$, each coupler
    has loss $e^{-\gamma}$ in each output port where $\gamma = 0.25$,
    and each edge of the strip obeys lossy boundary conditions $r_\pm
    = \exp(-0.2\pi)$, chosen to be roughly the same as in the
    experiment.  The coupling strength is $\theta = 0.4\pi$,
    approximately equal to the ``strong coupling'' (topologically
    nontrivial) configuration in the experiment.  The bandstructure
    for the lossless network ($\gamma = 0$ and $|r_\pm| = 1$) is
    plotted for comparison (red lines).  The lossy network exhibits
    edge states whose dispersion relations are almost identical to the
    lossless network's topological edge states, and have lower loss
    than all other states (blue dashes).}
  \label{fig:projected}
\end{figure}

In the experiment, $r_\pm$ is determined based on two parameters
$\phi$ and $k$, with all other parameters (such as the coupling
matrices) fixed; varying $k$ then gives results like those shown in
Fig.~\ref{fig:seigs} and Fig.~\ref{fig:winding}.  Alternatively, we
could either (i) determine $\phi$ given $r_\pm$ and $k$ (by finding
the eigenvalues of the scattering matrix for one period of an infinite
strip \cite{pasek}), or (ii) determine $k$ given $r_\pm$ and $\phi$
(by finding the eigenvalues of the transfer matrix for one period of
the strip). Both procedures yield the projected bandstructure, with
$r_\pm$ interpreted not as reflection coefficients, but as
\textit{boundary conditions} which are applied along the edge of the
strip, specifying the phase shifts on the edge links.  In the absence
of losses, if the scattering experiment yields a non-zero winding
number for $\phi$ in a bulk bandgap, then it can be seen that the
projected bandstructure exhibits topological edge states \cite{pasek}.

Now, consider a lossy network.  The losses can be described, without
loss of generality, by making the coupling matrices sub-unitary, as
well as by setting $|r_\pm| < 1$ (lossy boundary conditions).  The two
procedures described in the preceding paragraph, (i) and (ii), lead to
two distinct types of complex projected bandstructure; we focus on
procedure (ii), which takes $r_\pm$ and a real $\phi$ and yields
complex wave-numbers $k$, whose imaginary parts are the attenuation
constants of the propagating modes.  One such complex bandstructure is
shown in Fig.~\ref{fig:projected}, generated using representative
lossy $r_\pm$ and coupling matrices (a 5-cell-wide strip is used for
clarity).  The bandstructure contains clearly-defined edge states
whose dispersion curves have real parts nearly indistinguishable from
the lossless network's topological edge states.  Apart from these,
there also exist propagating bulk states.  Even in the range of $\phi$
corresponding to the lossless network's bulk bandgap, there are bulk
states with $\mathrm{Re}[k] \ne 0$, which are continuable to purely
evanescent ($\mathrm{Re}[k] = 0$) states of the lossless system.
However, these bulk states have significantly larger attenuation
$|\mathrm{Im}(k)|$ compared to the edge states, which can be
attributed to the fact that they experience losses over a wider area.
If we wind $\mathrm{arg}[r_\pm]$ through $2\pi$, with fixed $|r_\pm| <
1$ (corresponding to a sub-unitary trajectory in the complex plane),
the part of the complex projected bandstructure corresponding to the
bulk states remains nearly unchanged, in both its real and imaginary
parts.  For the edge states, the real part of the dispersion curve
winds with $\mathrm{arg}[r_\pm]$, whereas the imaginary part remains
nearly unchanged.  Hence, winding $\mathrm{arg}[r_\pm]$ has the effect
of ``pumping'' a branch of lossy edge states across one quasi-energy
period.

In summary, we have observed a topological edge invariant, based on
topological pumping, in a classical microwave network.  Since networks
of this sort are rather simple to set up, this may be a convenient
method for studying the physics of topological bandstructures.  For
instance, we have discussed how the quasi-energy bandstructures which
arise in networks can have the same properties as the bandstructures
of Floquet topological insulators
\cite{Oka,Inoue,Demler0,Demler,Lindner,Gu}.  In the square-lattice
network chosen for the present experiment, the topologically
nontrivial phase is an ``anomalous Floquet insulator'', which
possesses topological edge states despite all bands having zero Chern
number \cite{Demler0}.  In other network geometries, such as hexagonal
lattices, different parameter choices give rise to either anomalous
Floquet insulator phases, Chern insulator phases (in which the Chern
numbers are nonzero), or conventional insulator phases \cite{pasek}.
If couplers with tunable coupling strengths are implemented, the
transitions between these various phases could be directly observed.
In future work, it would be desirable to reduce the losses in the
network components, though it is encouraging that a non-zero winding
is already clearly observable even at the current loss levels.  Based
on our above arguments, the edge states are physically meaningful
despite the presence of loss; although both edge states and bulk
states undergo attenuation, the two types of state are clearly
distinguishable in the complex projected bandstructure.  Since the
edge states receive significantly less attenuation, they are the
dominant mode of transmission along the edge.

\section{Acknowledgments}

We are grateful to M.~Hafezi, M.~Rechtsman, B.~Zhang, C.~Soci, and
N.~Zheludev for helpful discussions.  This research was supported by
the Singapore National Research Foundation under grant
No.~NRFF2012-02, and by the Singapore MOE Academic Research Fund Tier
3 grant MOE2011-T3-1-005.

\appendix

\section{Scattering and transfer matrices}

\begin{figure}[b]
  \centering\includegraphics[width=0.27\textwidth]{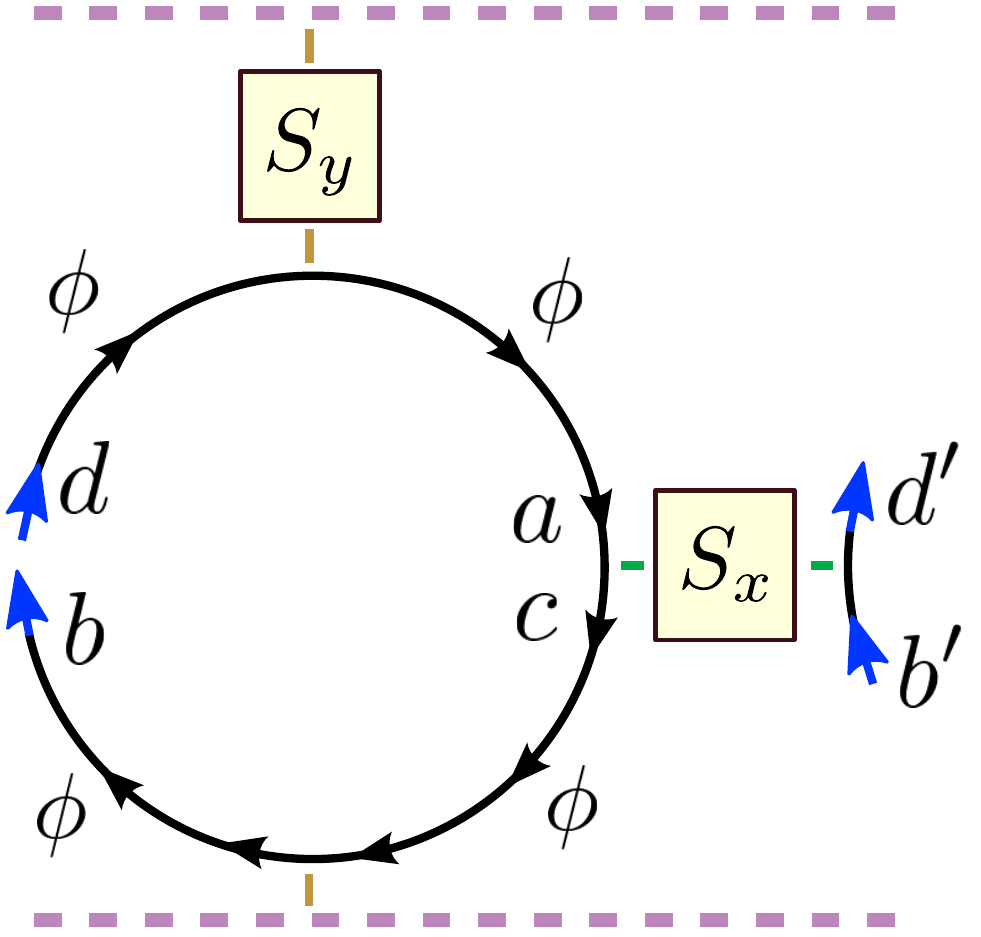}
  \caption{Schematic of one cell of the network.  The variables $a$,
    $b$, $c$, $d$, $d'$, and $b'$ label the complex wave amplitudes at
    the indicated points along the links (cables).  The boxes labeled
    $S_x$ and $S_y$ are $2\times2$ directional couplers.  The
    horizontal dashed lines indicate boundaries related by a twisted
    boundary condition with twist angle $k$.}
  \label{fig:couplings}
\end{figure}

Fig.~\ref{fig:couplings} shows one cell of the network in our pumping
experiment.  In terms of the cylindrical geometry of the topological
pumping experiment (as depicted in Fig.~1 of the main text), the
horizontal direction in this figure corresponds to the cylinder axis,
and the vertical direction corresponds to the cylinder's azimuthal
direction.  The inputs to the cell are the complex wave amplitudes
$\{d,b'\}$, and the outputs are $\{b,d'\}$.  We can increase the
separation between the left and right edges of the network by
``stacking'' identical cells along the horizontal (axial) direction.

We first consider the lossless case.  Within each cell, the wave
amplitudes are multiplied by $\exp(i\phi)$ when crossing each link.
At the couplers, the wave amplitudes are related by
\begin{align}
  S_y \begin{bmatrix}d\, e^{i\phi} \\ c\,e^{i(\phi+k)} \end{bmatrix} &=
  \begin{bmatrix}b\,e^{-i(\phi-k)} \\ a\,e^{-i\phi} \end{bmatrix}, \label{Sy}\\
  S_x \begin{bmatrix}a \\ b' \end{bmatrix} &=
  \begin{bmatrix}d' \\ c \end{bmatrix}, \label{Sx}
\end{align}
where $S_{x,y}$ are $2\times2$ unitary matrices.  We can combine
(\ref{Sy})--(\ref{Sx}) to eliminate $a$ and $c$; this results in a
transfer matrix equation of the form
\begin{equation}
  M(\phi, k) \begin{bmatrix}d \\ b \end{bmatrix} = \begin{bmatrix}d' \\ b' \end{bmatrix},
  \label{Meq}
\end{equation}
\begin{widetext}
  \noindent
  where
\begin{equation}
  M(\phi,k) \equiv \frac{1}{S_x^{22}S_y^{12}}
  \begin{bmatrix}
    -\det(S_x)\det(S_y)e^{2i\phi} - S_x^{12}S_y^{11}e^{-ik} &
    \det(S_x)S_y^{22}e^{ik} + S_x^{12}e^{-2i\phi} \\
    S_x^{21}\det(S_y)e^{2i\phi}-S_y^{11}e^{-ik} &
    -S_x^{21}S_y^{22}e^{ik} + e^{-2i\phi}
  \end{bmatrix}.
  \label{Mdef}
\end{equation}
The transfer matrix across one cell satisfies $\det(M) =
S_x^{11}S_y^{21}/S_x^{22}S_y^{12}$; in this sense, it is generally not
a ``reciprocal'' transfer matrix, unlike the transfer matrices
encountered in wave propagation problems.  That is to be expected,
since the network's input and output ports are inequivalent, and the
links in the network are directional.

Several different quantities can be computed from
Eqs.~(\ref{Meq})--(\ref{Mdef}).  Firstly, if the stack is infinite in
extent, we can look for Bloch solutions satisfying
\begin{equation}
  d' = e^{ik_x}\, d, \quad b' = e^{ik_x}\, b.
\end{equation}
Hence, the eigenvalues of $M(\phi,k)$ are values for $\exp(ik_x)$,
where $k_x$ is the quasimomentum in the $\hat{x}$ direction, for given
values of quasi-energy $\phi$ and $\hat{y}$-quasimomentum $k$.  This
is essentially the calculation for the bulk quasi-energy
bandstructure.  The solution can be obtained analytically
\cite{Liang}, by using the convenient parameterization
\begin{equation}
  S_\mu = 
  \begin{bmatrix}
    \sin(\theta_\mu)\,e^{i\chi_\mu} & -\cos(\theta_\mu)\,e^{i(\varphi_\mu-\xi_\mu)} \\
    \cos(\theta_\mu)\,e^{i\xi_\mu} & \sin(\theta_\mu)\,e^{i(\varphi_\mu-\chi_\mu)}
  \end{bmatrix}
  , \quad \mu = x,y.
\end{equation}
\end{widetext}
\noindent
The angle $\theta_\mu$ describes the ``coupling strength'', with
$\theta = 0$ corresponding to zero coupling between adjacent cells.
One can show, after tedious calculation, that the quasi-energy
bandstructure is gapless for $\theta_x + \theta_y = \pi/2$
\cite{Liang}.  This defines two distinct insulator phases: a
weak-coupling insulator ($\theta_x + \theta_y < \pi/2$) and a
strong-coupling insulator ($\theta_x + \theta_y > \pi/2$).  These will
turn out to be topologically distinct as well.

We now stack $n$ cells together to form a one-dimensional lattice.
Let $\{d_L,b_L,d_R,b_R\}$ be the input/output amplitudes on the left
and right edges of the stack.  Then
\begin{equation}
  M_n(\phi,k) \begin{bmatrix}d_L \\ b_L \end{bmatrix} = \begin{bmatrix}d_R \\ b_R \end{bmatrix}, \quad \mathrm{where} \;\; M_n \equiv \left[M(\phi, k)\right]^n.
\end{equation}
This can be rearranged into a scattering matrix equation,
\begin{equation}
  S_n(\phi,k) \begin{bmatrix}d_L \\ b_R \end{bmatrix} = \begin{bmatrix}b_L \\ d_R \end{bmatrix},
  \label{Sn}
\end{equation}
where
\begin{equation}
  S_n \equiv  \frac{1}{M_n^{22}}\,
  \begin{bmatrix}
    -M_n^{21} & 1 \\ \det(M_n) & M_n^{12}
  \end{bmatrix}.
\end{equation}
Without losses, $S_n(\phi,k)$ is unitary and its two eigenvalues are
unimodular.  The diagonal entries of $S_n$ are the left and right
reflection coefficients, and the off-diagonal entries are the
transmission coefficients.

\begin{figure}
  \centering\includegraphics[width=0.45\textwidth]{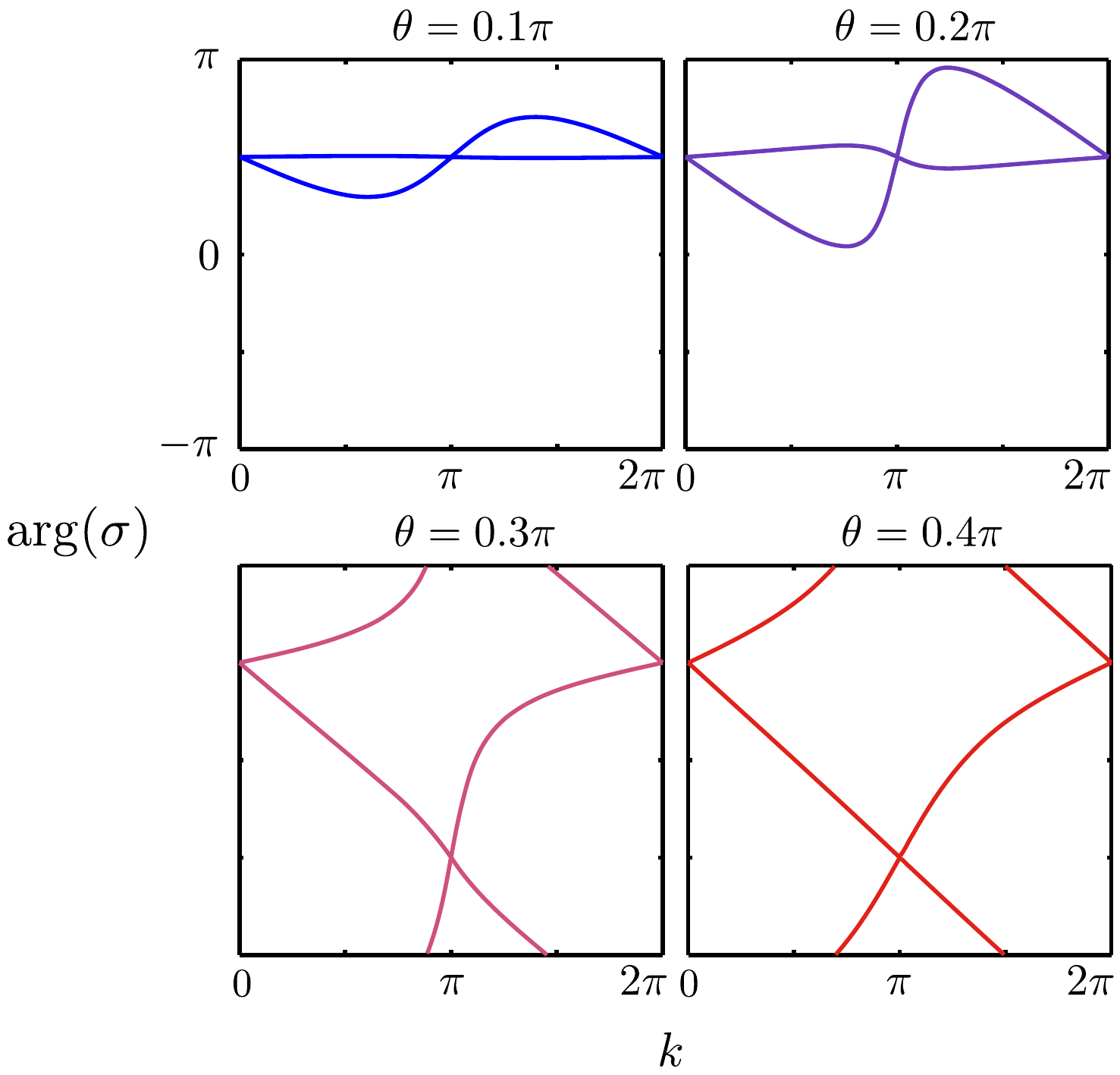}
  \caption{Complex argument of the $S_n$ matrix eigenvalues, $\sigma$,
    for lossless networks with $n = 10$ cells.  Four different values
    of the coupling strength $\theta$ are shown.  The $S_x$ and $S_y$
    coupling matrices are taken to be equal and lossless with $\varphi
    = \chi = 0$ and $\xi = \pi/2$.  The quasi-energy is $\phi =
    0.25\pi$ (mid-gap).  The system is topologically trivial for
    $\theta < \pi/4$, and non-trivial for $\theta > \pi/4$.}
  \label{fig:winding_ideal}
\end{figure}

We are interested in the variation of the eigenvalues of $S_n$ with
$k$, and specifically their winding numbers.  If $\phi$ is chosen to
be in the bulk bandgap, then for large $n$ (i.e., a long cylinder),
the transmission coefficients go to zero and $S_n$ reduces to a
diagonal matrix.  Then the two eigenvalues are simply the two
reflection coefficients, and are predicted to have winding number 0 in
the weak-coupling insulator phase, and winding numbers $\pm 1$ in the
strong-coupling insulator phase.  The numerical results agree, as
shown in Fig.~\ref{fig:winding_ideal}.  This winding behavior can be
deduced from the principle that bound states induce $\pi$ phase shifts
in the reflection coefficient \cite{Brouwer}, or by considering how
edge state branches are transported across the bandgap in the
projected quasi-energy bandstructure \cite{pasek}.

It is instructive to see how this works in the limits of very weak or
very strong coupling.  Firstly, for $\theta_x = \theta_y = 0$, the
reflection coefficients are $k$-independent:
\begin{equation}
  \frac{b_L}{d_L} = - e^{i(4\phi+\xi_x+\varphi_y)}, \quad
  \frac{d_R}{b_R} = - e^{i(\varphi_x - \xi_x)}.
\end{equation}
And for $\theta_x = \theta_y = \pi/2$,
\begin{equation}
  \frac{b_L}{d_L} = e^{i(2\phi + \chi_y)} \, e^{-ik}, \quad
  \frac{d_R}{b_R} = e^{i(2\phi+\varphi_x+\varphi_y-\chi_y)} \, e^{ik},
\end{equation}
which results in winding numbers of $\pm 1$.  For these two limiting
cases, the bulk bands are completely flat.  For intermediate values of
$\theta$, however, it is necessary to choose a value of $\phi$ that
lies in the bulk bandgap, for the topological pumping procedure to
make sense.

\section{Losses}

Losses can be incorporated into the model by multiplying each entry of
the coupling matrices, $S_x$ and $S_y$, by damping factors.  This
makes the coupling matrices sub-unitary:
\begin{equation}
  S_\mu = 
  \begin{bmatrix}
    \sin(\theta_\mu)\,e^{i\chi_\mu} e^{-\gamma_\mu^{11}} & -\cos(\theta_\mu)\,e^{i(\varphi_\mu-\xi_\mu)} e^{-\gamma_\mu^{12}}\\
    \cos(\theta_\mu)\,e^{i\xi_\mu} e^{-\gamma_\mu^{21}}& \sin(\theta_\mu)\,e^{i(\varphi_\mu-\chi_\mu)} e^{-\gamma_\mu^{22}}
  \end{bmatrix},
\end{equation}
where $\mu = x,y$.  In principle, each of the $\exp(-\gamma_\mu^{ij})$
factors can be independent.  Losses occurring within the cables
(links) or phase shifters can be incorporated into these coupling
matrix damping factors, without loss of generality.

\begin{figure}
  \centering\includegraphics[width=0.39\textwidth]{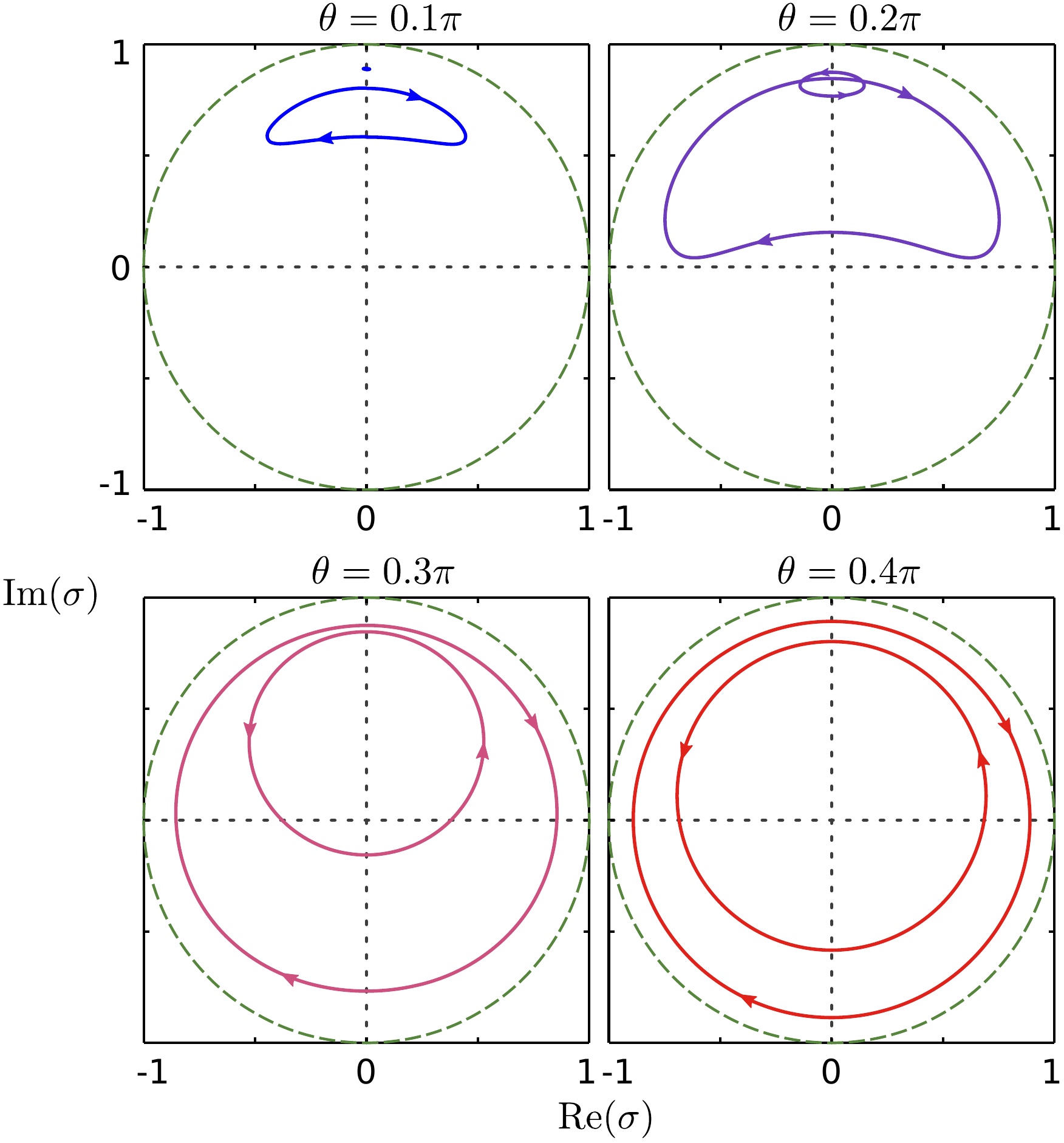}
  \caption{Trajectories of the $S_n$ matrix eigenvalues in the complex
    plane, as $k$ advances from 0 to $2\pi$.  The network parameters
    are the same as in Fig.~\ref{fig:winding_ideal}, except that loss
    is present: each coupling matrix $S_x$ and $S_y$ is multiplied by
    a uniform loss factor of $e^{-\gamma} = 0.9$.}
  \label{fig:theory_seigs}
\end{figure}

When losses are present, the $S_n$ matrix defined in Eq.~(\ref{Sn})
becomes sub-unitary, i.e.~its eigenvalues lie inside the unit circle.
Hence, in the large-$n$ limit the reflection coefficients have
magnitudes smaller than unity, reflecting the fact that the input
waves undergo dissipation within the network.

Fig.~\ref{fig:theory_seigs} shows the trajectories of the eigenvalues
of $S_n$ in the complex plane, as $k$ is wound through $2\pi$, for a
lossy network.  Here, we have taken a uniform factor
$\exp(-\gamma_\mu^{ij}) = 0.9$ in all couplers.  It can be seen that
the behavior is similar to our experimental results (Fig.~3 of the
main text).  In particular, deep in the topologically trivial or
non-trivial phase ($\theta = 0.1\pi$ and $\theta = 0.4\pi$
respectively), the eigenvalue trajectories are clearly continuable to
the zero-winding and nonzero-winding behavior of the lossless system.

\section{Insulation and edge state localization}

\begin{figure}
  \centering\includegraphics[width=0.335\textwidth]{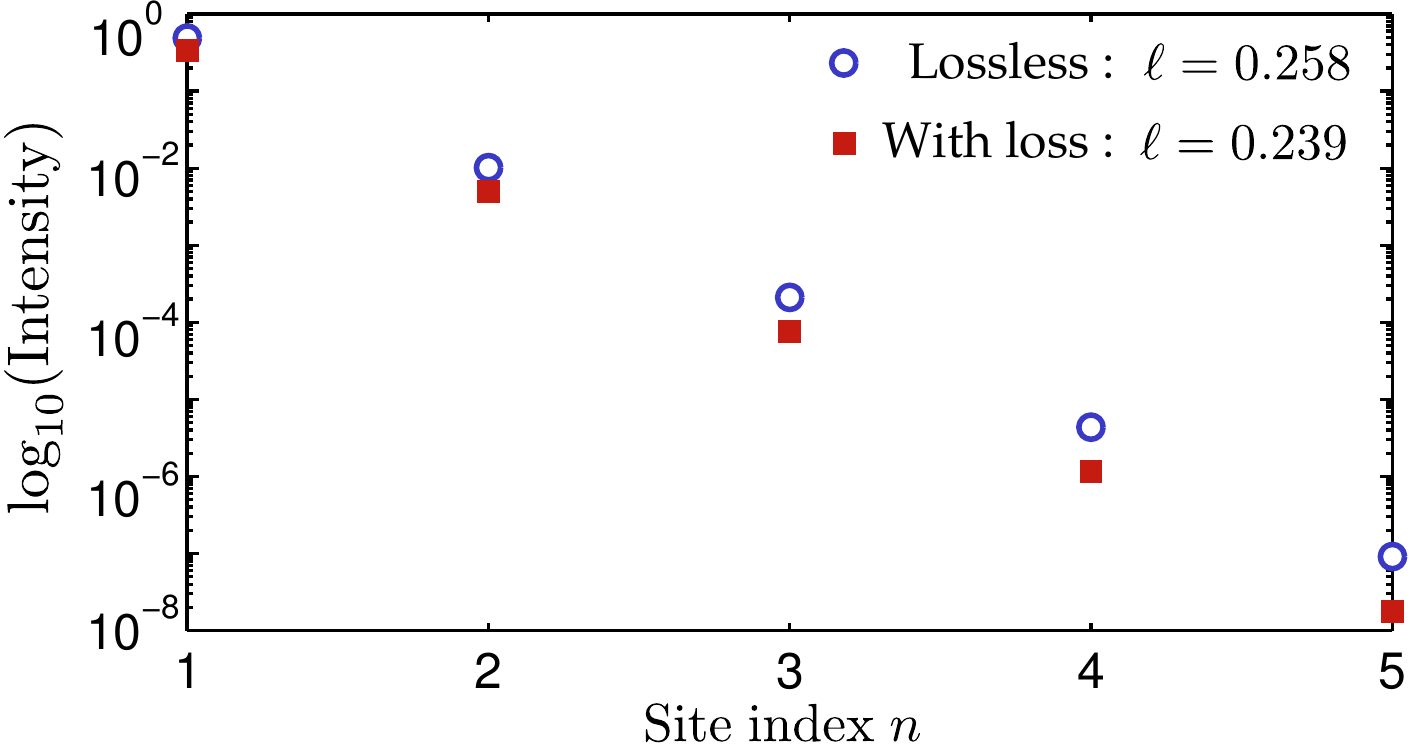}
  \caption{Semilog plot of the edge state intensity versus ring index
    $n$, for lossless (blue circles) and lossy (red squares) networks
    respectively. The network consists of an infinite strip 5 unit
    cells wide, with line delay $\phi = 0.2\pi$ and coupler strength
    $\theta = 0.4\pi$ (which matches the ``strong coupling'' regime of
    the experiment). For the lossy system, we add loss $e^{-\gamma}$
    to each quarter-ring segment, where $\gamma = 0.25$. For both
    cases, the edge state at $k = -0.4\pi$ is shown. The intensity at
    each unit cell is defined as the sum of the absolute squares of
    the wave amplitudes on the four quarter-ring segments
    (cf.~Fig.~\ref{fig:couplings}).}
  \label{fig:penetration_depth}
\end{figure}

Strictly speaking, topologically nontrivial behavior emerges in a
sample only in the limit where opposite edges are spatially
well-separated.  In terms of the topological edge states, the
wavefunctions for edge states on opposite edges must have negligible
overlap.

In order to determine the penetration depth of the edge states, we
consider a hypothetical strip with an ample width of 5 unit cells in
the $\hat{x}$ direction (and infinite in the $\hat{y}$ direction).
Fig.~\ref{fig:penetration_depth} shows the semilog plot of the
intensity (defined as the absolute square of the wave amplitudes in
the links) versus distance along the $\hat{x}$ direction (measured in
lattice units), for a typical mid-gap edge state.  Two distinct cases
are considered: with loss levels similar to those in the experiment,
and without loss.  All other network parameters, including the
coupling constant $\theta$, are chosen to be similar to the
topologically nontrivial regime of the actual experiment.

\begin{figure}
  \centering\includegraphics[width=0.45\textwidth]{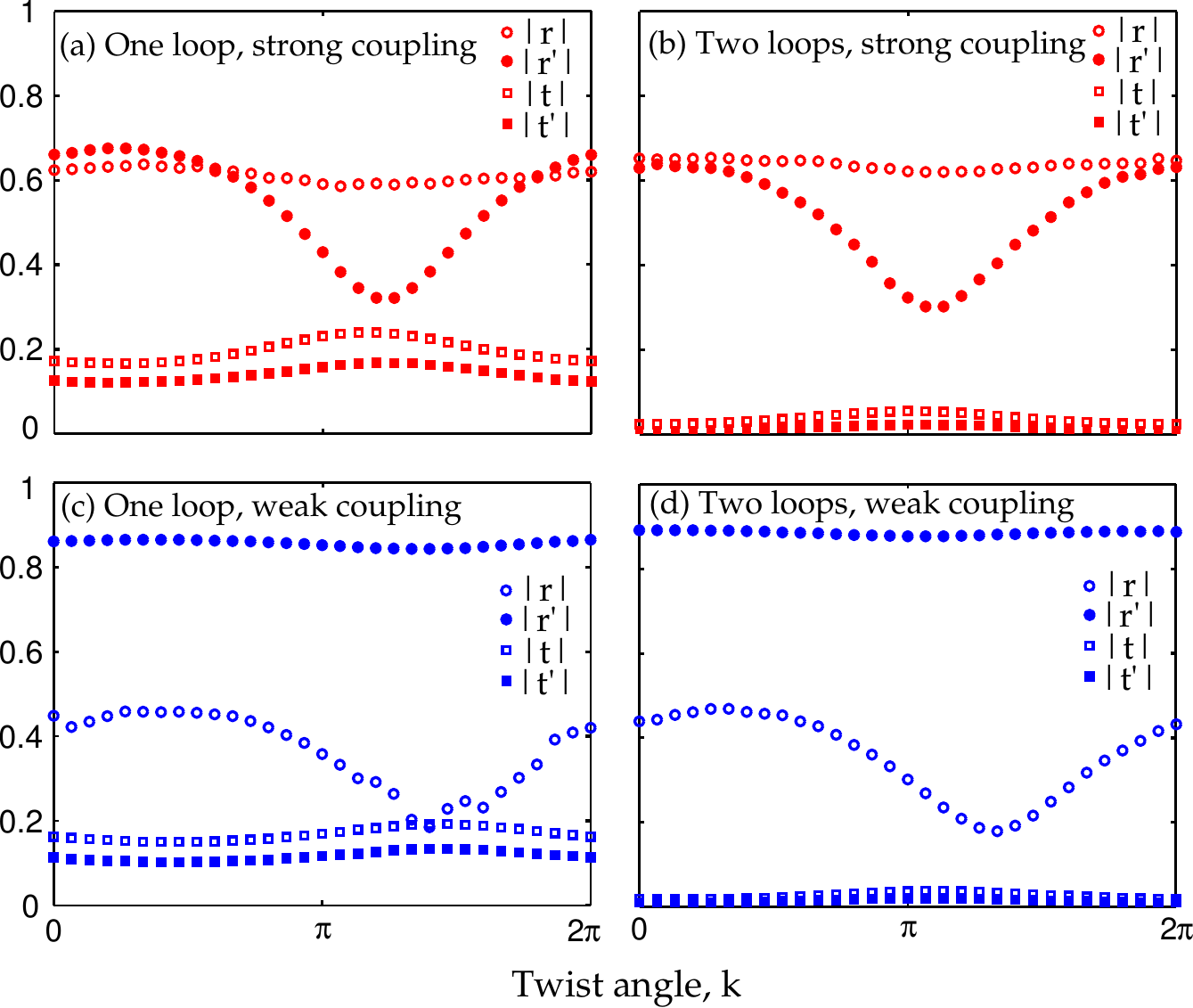}
  \caption{Experimentally-measured reflection coefficient magnitudes
    ($|r|$ and $|r'|$, circle symbols) and transmission coefficient
    magnitudes ($|t|$ and $|t'|$, square symbols), versus phase shift
    $k$.  Results are shown for the strong and weak coupling regimes,
    and for the one-loop and two-loop configurations.  Unprimed and
    primed coefficients refer to waves incident on the left and right
    edges, respectively.  In the two-loop configurations, the
    transmission becomes negligible.  }
  \label{fig:t_r_coeff}
\end{figure}

From this plot, we see that the edge states in both the lossless and
lossy networks are exponentially localized to one edge: the intensity
scales as $I_n \propto \exp(-n/\ell)$, where $n$ is the site index in
the $\hat{x}$ direction.  A numerical fit gives a penetration depth of
$\ell \approx 0.258$ unit cells for the lossless network.  Introducing
losses decreases this only slightly, to $\ell \approx 0.239$ unit
cells.  Hence, we can conclude that the edge states have negligible
overlap even for a network two unit cells wide, as in the experiment.
Furthermore, the penetration depth is mainly determined by the
bandgap, not by the losses in the network.

We can also verify that the edges are spatially well-separated by
using direct experimental data.  In the scattering formulation of the
topological pump (described in Section II.B of the main text), the
edges are well-separated if waves incident on one edge of the
cylindrical sample have negligible transmission to the opposite edge
\cite{Brouwer}.  Fig.~\ref{fig:t_r_coeff} shows reflection and
transmission coefficients measured by the network analyzer in the
experiment.  For both strong and weak couplings, the transmission
decreases significantly when going from the single-loop configuration
(corresponding to a one cell wide cylinder) to the two-loop
configuration (corresponding to a two cell wide cylinder).  For the
latter, the magnitude of the transmission coefficient is less than
0.06 (for strong coupling) and less than 0.04 (for weak coupling).
Thus, this line of reasoning likewise indicates that we are justified
in interpreting the behavior of the two-loop network in terms of
topological insulator physics.


\begin{thebibliography}{99}
\bibitem{MStone} M.~Stone, \textit{Quantum Hall Effect} (World
  Scientific, 1992).
\bibitem{Moore} J.~E.~Moore, Nature 464, 194 (2010).
\bibitem{Raghu1} F.~D.~M.~Haldane and S.~Raghu, Phys.~Rev.~Lett.~{\bf
  100}, 013904 (2008).
\bibitem{Raghu2} S.~Raghu and F.~D.~M.~Haldane, Phys.~Rev.~A {\bf 78},
  033834 (2008).
\bibitem{Wang1} Z.~Wang, Y.~D.~Chong, J.~D.~Joannopoulos, and
  M.~Solja\u{c}i\'{c}, Phys.~Rev.~Lett.~{\bf 100}, 013905 (2008).
\bibitem{Wang2} Z.~Wang, Y.~D.~Chong, J.~D.~Joannopoulos, and
  M.~Solja\u{c}i\'{c}, Nature {\bf 461}, 772 (2009).
\bibitem{WenjieChen} W.-J.~Chen, S.-J.~Jiang, X.-D.~Chen, J.-W.~Dong,
  and C.~T.~Chan, arXiv:1401.0367.
\bibitem{Rechtsman} M.~C.~Rechtsman, J.~M.~Zeuner, Y.~Plotnik, Y.~Lumer, D.~Podolsky, F.~Dreisow, S.~Nolte, M.~Segev, and A.~Szameit, Nature {\bf 496}, 196 (2013).
\bibitem{hafezi} M.~Hafezi, E.~A.~Demler, M.~D.~Lukin, and
  J.~M.~Taylor, Nature Phys.~{\bf 7}, 907 (2011).
\bibitem{hafezi2} M.~Hafezi, S.~Mittal, J.~Fan, A.~Migdall, and J.~M.~Taylor, Nat.~Photonics {\bf 7}, 1001 (2013).
\bibitem{Fan} K.~Fang, Z.~Yu, and S.~Fan, Nature Phot.~{\bf 6},
  782 (2012).
\bibitem{LeHur}  J.~Koch, A.~A.~Houck, K.~Le~Hur, and S.~M.~Girvin, Phys.~Rev.~A {\bf 82}, 043811 (2010).
\bibitem{LeHur1} A.~Petrescu, A.~A.~Houck, and K.~Le~Hur, Phys.~Rev.~A {\bf 86}, 053804 (2012).
\bibitem{Khanikaev} A.~B.~Khanikaev, S.~H.~Mousavi, W.-K.~Tse,
  M.~Kargarian, A.~H.~MacDonald, and G.~Shvets, Nature Materials {\bf
    12}, 233 (2013).
\bibitem{solitons} Y.~Lumer, Y.~Plotnik, M.~C.~Rechtsman, and
  M.~Segev, Phys.~Rev.~Lett.~{\bf 111}, 243905 (2013).
\bibitem{klitzing} K.~v.~Klitzing, G.~Dorda, and M.~Pepper,
  Phys.~Rev.~Lett.~{\bf 45}, 494 (1980).
\bibitem{TKNN} D.~J.~Thouless, M.~Kohmoto, M.~P.~Nightingale, and M.~den
  Nijs, Phys.~Rev.~Lett.~{\bf 49}, 405 (1982).
\bibitem{Laughlin} R.~B.~Laughlin, Phys.~Rev.~B {\bf 23}, 5632 (1981).
\bibitem{Halperin} B.~I.~Halperin, Phys.~Rev.~B {\bf 25}, 2185 (1982).
\bibitem{Brouwer0} P.~W.~Brouwer, Phys.~Rev.~B {\bf 58}, R10135 (1998).
\bibitem{Brouwer} D.~Meidan, T.~Micklitz, and P.~W.~Brouwer,
  Phys.~Rev.~B {\bf 84}, 195410 (2011).
\bibitem{Fulga} I.~C.~Fulga, F.~Hassler, and A.~R.~Akhmerov,
  Phys.~Rev.~B {\bf 85}, 165409 (2012).
\bibitem{pasek} M.~Pasek and Y.~D.~Chong, Phys.~Rev.~B {\bf 89},
  075113 (2014).
\bibitem{kraus} Y.~E.~Kraus, Y.~Lahini, Z.~Ringel, M.~Verbin, and
  O.~Zilberberg, Phys.~Rev.~Lett.~\textbf{109}, 106402 (2012).
\bibitem{verbin} M.~Verbin, O.~Zilberberg, Y.~Lahini, Y.~E.~Kraus, and
  Y.~Silberberg, arXiv:1403.7124.
\bibitem{hafezi3} M.~Hafezi, Phys. Rev. Lett. {\bf 112}, 210405 (2014).
\bibitem{Oka} T.~Oka and H.~Aoki, Phys.~Rev.~B {\bf 79}, 081406
  (2009).
\bibitem{Inoue} J.~I.~Inoue and A.~Tanaka, Phys.~Rev.~Lett.~{\bf 105},
  017401 (2010).
\bibitem{Demler0} T.~Kitagawa, M.~S.~Rudner, E.~Berg, and E.~Demler,
  Phys.~Rev.~A {\bf 82}, 033429 (2010)
\bibitem{Demler} T.~Kitagawa, E.~Berg, M.~Rudner, and E.~Demler,
  Phys.~Rev.~B {\bf 82}, 235114 (2010).
\bibitem {Lindner} N.~H.~Lindner, G.~Refael and V.~Galitski, Nature Physics {\bf 7}, 490-495 (2011).
\bibitem{Gu} Z.~Gu, H.~A.~Fertig, D.~P.~Arovas, and A.~Auerbach, Phys.~Rev.~Lett.~{\bf 107}, 216601 (2011).
\bibitem{Levin} M.~S.~Rudner, N.~H.~Lindner, E.~Berg, and M.~Levin,
  Phys.~Rev.~X {\bf 3}, 031005 (2013).
\bibitem{ChalkerCo} J.~T.~Chalker, and P.~D.~Coddington, J.~Phys.~C {\bf 21}, 2665 (1988).
\bibitem {Kramer} B.~Kramer, T.~Ohtsuki, and S.~Kettemann, Phys.~Rep. {\bf 417}, 211 (2005).
\bibitem{ryu} S.~Ryu, C.~Mudry, H.~Obuse, and A.~Furusaki, New J. Phys. {\bf 12}, 065005 (2010).
\bibitem{Liang} G.~Q.~Liang and Y.~D.~Chong, Phys.~Rev.~Lett.~{\bf
  110}, 203904 (2013).
\bibitem{Liang2} G.~Q.~Liang and Y.~D.~Chong, Int.~J.~Mod.~Phys.~B
  {\bf 28}, 1441007 (2014).
  \bibitem{Carusotto} T.~Ozawa and I.~Carusotto, Phys.~Rev.~Lett.~{\bf 112}, 133902 (2014).
  \bibitem{Bardyn} C.~-E.~Bardyn, S.~D.~Huber, and O.~Zilberberg, arXiv:1312.6894 (2013).
  \bibitem{Zeuner} J.~Zeuner, M.~C.~Rechtsman, Y.~Plotnik, Y.~Lumer, M.~S.~Rudner, M.~Segev, A.~Szameit, arXiv:1408.2191 (2014).
\end{thebibliography}
\end{document}